\documentstyle[11pt,multicol,epsfig]{article}
\twocolumn
\begin{document}  
\scriptsize

\renewcommand{\thefootnote}{\fnsymbol{footnote}}
\sloppy
\newcommand{\rp}{\right)}
\newcommand{\lp}{\left(}
\newcommand \be  {\begin{equation}}
\newcommand \bea {\begin{eqnarray}}
\newcommand \ee  {\end{equation}}
\newcommand \eea {\end{eqnarray}}

\input epsf

\title{ Three-state Neural Network:
from Mutual Information to the Hamiltonian }

\author{David R. Dominguez Carreta\\
ESCET, Universidad Rey Juan Carlos,\\
C.Tulip\'an, Mostoles, 28933 Madrid, Spain, \\
and
Elka Korutcheva
{\thanks{Permanent Address: G.Nadjakov Inst. Solid State Physics, 
Bulgarian Academy of Sciences, 1784 Sofia, Bulgaria }}\\
Dep. F\'{\i}sica Fundamental,\\ 
Universidad Nacional de Educaci\'on a Distancia,\\
c/Senda del Rey No9,\\
28080 Madrid, Spain }

\date{\today}
\maketitle


\begin{abstract}

The mutual information, $I$, 
of the three-state neural network can be obtained exactly 
for the mean-field architecture,
as a function of three macroscopic parameters:
the overlap, the neural activity and the {\em activity-overlap},
$i.e.$, the overlap restricted to the active neurons.
We perform an expansion of $I$ on the overlap and the 
activity-overlap,
around their values for neurons almost independent on the patterns.
From this expansion we obtain an expression for a Hamiltonian
which optimizes the retrieval properties of this system. 
This Hamiltonian has the form of a disordered 
Blume-Emery-Griffiths model.
The dynamics corresponding to this Hamiltonian is found.
As a special characteristic of such network, 
we see that information can survive even if no overlap is present.
Hence the basin of attraction of the patterns and the 
retrieval capacity is much larger than for the Hopfield network.
The extreme diluted version is analized,
the curves of information are plotted 
and the phase diagrams are built.


\end{abstract}



\section{Introduction}

The collective properties of neural networks,
such as the storage capacity and the overlap with
the memorized patterns,
have been a subject of intensive research in the last 
decade\cite{Pe92},\cite{HK91}.
However,
more precise measures of their performance
as an associative memory,
as the information capacity and the basins of attraction
of their retrieval states,
have received comparatively less 
attention\cite{Am89}-\cite{PA89}.
For some models as the sparse-code 
networks\cite{AG87}-\cite{Ok96},
or the three-state networks\cite{Ye89}-\cite{BV94},
where the patterns are not uniformly distributed,
an information-theoretical approach\cite{Sh48}-\cite{SK99} seems crucial.

Calculations of the Shannon mutual information ($I$) 
for the sparse-code network were made\cite{Pa80}-\cite{SS96}.
For low storage of patterns,
a few time steps are need to retrieve
them\cite{Am89},\cite{NT90}.
However, for large storage, 
only $im$perfect retrieval is possible.
The closer to saturation,
the larger the time steps required to dynamical retrieval.
So, first-time retrieval is not enough 
and it is interesting to study the information capacity
of recurrent networks.
To improve $I$ for this recurrent network, a scheme, 
based on a self-control threshold mechanism, 
was proposed\cite{DB98}.
This Self-Control Neural Network (SCNN) is an adaptive
scheme induced by the dynamics itself instead of imposing 
any external constraint on the activity of the neurons.
Such procedure successfully increases both $I$ and the 
basins of attraction of the patterns.
Similar mechanisms can improve $I$ for 
three-state low-activity networks\cite{BD98},
with diluted and fully-connected architectures.

Here we propose a new method,
based on direct use of the $I$ calculated in the 
mean-field approximation,
to obtain a Hamiltonian which maximizes $I$ within a large
range of values for the activity of the network.

A three-state neural network is defined by the use
of a set of $\mu=1,...,p$ $ternary$ patterns,
$\{\xi^{\mu}_{i} \in\{0,\pm 1\},\,\,i=1,...,N\}$,
which are independent random variables given 
by the probability distribution

\begin{equation}
p(\xi^{\mu}_{i})=
a\delta(|\xi^{\mu}_{i}|^{2}-1)
+(1-a)\delta(\xi^{\mu}_{i}) ,
\label{1.px}
\end{equation}
where $a$ is the $activity$ of the patterns
($\xi^{\mu}_{i}=0$ are the inactive states).
A low-activity three-state neural network corresponds
to the case where the distribution is not uniform,
$ie$, $a<2/3$.
In the limit $a=1$ the binary Hopfield model is
reproduced.

The information enclosed in a simple unit $\xi^{\mu}_{i}$
is given by the entropy of its probability,

\begin{equation}
H_{\mu i}=-a\ln(a/2)-(1-a)\ln(1-a).
\label{1.Hm}
\end{equation} 
One can define as sparse a code whose fraction 
of active neurons is very small and tends to zero in 
the thermodynamic limit\cite{Am89}. 
Sparse-code binary patterns can have a large load rate 
$\alpha\sim [a|ln(a)|]^{-1}$, 
where $\alpha$ is the ratio between the number of patterns $p$ 
and number of connection per neuron $N$ \cite{Am89},\cite{Pa80}.
However, 
the information per unit for the sparse code is 
$H_{\mu i}\sim a|\ln(a)|\ll 1$,
and it is not clear if the total information per connection,
$i=\sum_{i\mu} H_{\mu i}/\#\{J_{ij}\}=\alpha H_{\mu i}$
of such network is larger than the uniform 
(non-sparse) one.

Although ternary patterns with low-activity have
not been studied in the same proportion, 
they present a similar behavior\cite{Ye89}. 
Besides of the fact that ternary patterns are a step
towards an analog neural model,
they have the advantage that they can be generated 
with a bias but keeping their symmetric distribution
(both $\pm 1$ states are considered active).
An important question related to the three-state model 
is the measurement of the retrieval quality in the cases 
where this is imperfect.
The overlap alone is not anymore a good measure because
it accounts only for the active states.
For the homogeneous ternary patterns, 
the Hamming distance can be considered a good measure,
since it takes equally into account all errors in retrieving,
the active and the inactive one.
For the low-activity case, however,
also the Hamming distance is not a good parameter
of the retrieval quality because the errors in 
retrieving the active states are much more relevant
(they contain much more information) 
than the errors in retrieving the inactive states.
To solve this problem, 
we use the conditional probability
of neuron states given the pattern states\cite{BD98},
to obtain the mutual information $I$ of the 
attractor neural network (ANN).
This quantity measures directly the amount of dependence
between the random variables,
neurons and patterns.
To accomplish that, we must use a new variable,
we call $activity$-$overlap$,
which is the overlap between 
the active states of the ternary neurons and
the active states of the ternary patterns,
taken with the absolute value.

This $I$ is thus a function of three parameters:
the overlap $m$, 
the neural-activity $q$,
and the activity-overlap $n$.
We then expand the $I$ around the values of the
parameters when the neurons are independent on the patterns.
This expansion gives us an expression that can be 
interpreted as a Hamiltonian,
a function only of the neuron states and the synaptic couplings.
This Hamiltonian is similar to the 
Blume-Emery-Griffiths\cite{BE71}-\cite{KE99}
spin-1 model (BEG),
but with random interactions. 
The BEG model,
originally proposed to study $He_3-He_4$ mixtures,
was latter used to describe several systems,
like memory alloys, 
fluid mixtures, 
micro-emulsions, etc.,
and displays a variety of new thermodynamic phases.

Some disordered BEG models
have been recently studied \cite{SN97}-\cite{Br99},
where either the exchange-interactions or the
crystal-field are random variables.
However, from our knowledge,
no random biquadratic-interactions 
model has been treated up to this date.

We describe our model
in Section II.
In Section III we describe the $I$ measures used to
evaluate the performance of the ANN,
and derive the BEG Hamiltonian from $I$.
After solving the thermodynamics for this model 
in Section IV,
we present some results for the dynamics and the
phase diagrams in the section V,
comparing the results with previous works.
We conclude in the last Section with some comments 
about possible improvements of the network.

\section{The model}

As well as the pattern states,
the neuron states at time $t$ are three-state variables,
defined as 

\begin{eqnarray}
\sigma_{it}\in\{0,\pm 1\},\,i=1,...,N.
\label{2.si}
\end{eqnarray}
They are updated according to a stochastic dynamics
which depends on the previous states $\{\sigma_{i,t-1}\}$
and on synaptic interactions between different neurons.
The specific form of the synapses will be obtained latter,
by construction.
We will see they are of the Hebbian type,
that is, the learning is local 
(the synapses depend only on the two neurons interacting).
Moreover, 
the updating rule will be also obtained by construction,
no supposition being done here except that the patterns 
have the same three-state symmetry as the neuron states.

The three-state patterns 
$\xi^{\mu}_{i} \in\{0,\pm 1\}$, $\mu=1,...,p$, 
are independent identically distributed random variables
($IIDRV$) chosen according to the probability 
distribution in Eq.(\ref{1.px}).
There is no bias ($\langle\xi^{\mu}_{i}\rangle=0$)
neither correlation between patterns 
($\langle\xi^{\mu}_{i}\xi^{\nu}_{i}\rangle=0$),
and $a=\langle|\xi^{\mu}_{i}|^{2}\rangle$ is the 
activity of the patterns.

The mean-field networks have the property of 
being site-independent, 
that means,
the correlations between different sites are negligeable
in the thermodynamic limit, $N\to\infty$.
This implies that every macroscopic quantity satisfies the 
conditions of the law of large numbers ($LLN$),
so they can be defined as an average on the probability
distribution of a state in a single site.  
If $f$ is the thermodynamic limit of the variable $f_{N}$,
we have

\begin{equation}
F_{N}\equiv
{1\over N}\sum_{i}f_{i}(\sigma_{i},\xi_{i})
\to F=<f(\sigma,\xi)>_{\sigma,\xi},
\label{2.FN}
\end{equation}
where the brackets represent averages over the distribution of a
single, typical state $\sigma,\xi$ 
(we can drop the index $i$).

An example of this is the special case of the overlap,
but this property is valid also for every function $F_{N}$,
since it comes from a property of the probability distribution 
of the states itself,

\begin{equation}
p(\{\sigma\},\{\xi\})=
\prod_{i} p(\sigma_{i},\xi_{i}) .
\label{2.ps}
\end{equation}

The task of retrieval is successful 
if the distance between the state of the neuron 
$\{\sigma_{it}\}$ and the pattern $\{\xi^{\mu}_i\}$, 
defined as

\begin{eqnarray}
d^{\mu}_{t}&\equiv& 
{1\over N}\sum_{i}|\xi^{\mu}_{i}-\sigma_{it}|^{2}
\nonumber\\
&=& a-2am^{\mu}_{Nt}+q_{Nt} ,
\label{2.Em}
\end{eqnarray}
becomes small after some time $t$.
This is the so-called Hamming distance
(an Euclidean quadratic measure for discrete sets).
The overlap of the $\mu$th pattern with the neuron-state
is defined as: 

\begin{equation}
m^{\mu}_{Nt}\equiv
{1\over aN}\sum_{i}\xi^{\mu}_{i}\sigma_{it},
\label{2.mm}
\end{equation}
while the neural activity is

\begin{equation}
q_{Nt}\equiv
{1\over N}\sum_{i}|\sigma_{it}|^{2}.
\label{2.qN}
\end{equation}
The $m^{\mu}_{Nt}$ are called the
$retrieval$ $overlaps$, and
they are normalized parameters within the 
interval $[-1,1]$, 
which attain the extreme values
$m^{\mu}_{Nt}=\pm1$ whenever 
$\sigma_{j}=\pm\xi^{\mu}_{j}$, 
as we see from  Eq.(\ref{1.px}).

Another parameter is need to 
define completely the macroscopic state of the ANN.
This is

\begin{equation}
n^{\mu}_{Nt}\equiv 
{1\over aN}\sum_{i}^{N}|\sigma_{it}|^{2}|\xi^{\mu}_{i}|^{2}.
\label{2.nN}
\end{equation}

We call this quantity the $activity$-$overlap$\cite{BD98},
as long as $n^{\mu}_{Nt}$ represents the overlap between
the sites, where the neurons are active, 
$|\sigma_{it}|=1$, and 
the sites where the patterns are active,
$|\xi_{i}^{\mu}|=1$.
For the dynamics used in most work found in the 
literature\cite{Ye89},\cite{BV94},\cite{BS94},
where the synapses used are of the Hopfield form
$J_{ij}=\sum_{\mu}\xi_{i}^{\mu}\xi_{j}^{\mu}$,
this parameter $n^{\mu}_{Nt}$ does not seem to play 
any role in the evolution of the network,
independent of the architecture considered 
(diluted, layered or fully-connected for instance). 
However $n^{\mu}_{Nt}$ is necessary to define the 
mutual information of the network, 
as well as this is necessary in computing the network's 
performance \cite{BR94},\cite{SW97}.

For this Hopfield three-state network a self-control (SC)
mechanism was recently introduced with the following
threshold dynamics \cite{BD98}:
$\theta_{t}=c(a)\Delta_{t}$,
where $c(a)=\sqrt{-2\ln(a)}$ is a function only of the 
pattern activity,
while the variance of the cross-talk noise
(due to the $p-1$ non-retrieved patterns)
has the simple form
$\Delta_{t}=\sqrt{\alpha q_{Nt}}$ for
the diluted architecture\cite{DG87}.
Here we take the alternative approach of starting from the
mutual information for the model,
which we describe in the next Section.

\section{Mutual Information}

\subsection{Mean-Field-Theory}

Compared to the binary neural network (NN), 
where the natural parameter is the overlap $m^{\mu}$,
to describe the statistical macro-dynamics of 
the three-state NN, 
there should be two additional parameters.
Although the only variables appearing in the usual
Hopfield dynamics are the overlap $m^{\mu}_{Nt}$ 
and the neural activity $q_{Nt}$,
the activity-overlap $n^{\mu}_{Nt}$ is an independent 
parameter which complete the macroscopic description.
For a long-range system, as the one we are considering,
it is enough to observe the distribution of a single
typical neuron in order to know the global distribution.

The conditional probability of having a neuron 
in a state $\sigma_{it}=\sigma$ in a time $t$,
given that in the same site the pattern being retrieved
is $\xi^{\mu}_{i}=\xi$, is:
 
\begin{eqnarray}
&&p(\sigma|\xi)=
(s_{\xi}+m\xi\sigma)\delta(\sigma^{2}-1)+
(1-s_{\xi})\delta(\sigma),\nonumber\\
&&s_{\xi}\equiv s+{n-q\over 1-a}\xi^{2},\,\,
s\equiv {q-na\over 1-a}.
\label{3.ps}
\end{eqnarray}
One can verify that this probability satisfies the
averages:

\begin{eqnarray}
&&m= \langle\langle
\sigma\rangle_{\sigma|\xi}{\xi\over a}\rangle_{\xi},
    \nonumber\\
&&q= \langle\langle
\sigma^{2}\rangle_{\sigma|\xi}\rangle_{\xi},
    \nonumber\\
&&n= \langle\langle
\sigma^{2}\rangle_{\sigma|\xi}{\xi^{2}\over a}\rangle_{\xi}.
\label{3.ma}
\end{eqnarray}
These are the thermodynamic limits ($N\to\infty$) of 
Eqs.(\ref{2.mm},\ref{2.qN},\ref{2.nN}),
for a given time $t$ and pattern $\mu$.
Due to the symmetry of the patterns, 
we have also 
$\langle\langle\sigma\rangle_{\sigma|\xi}\rangle_{\xi}
=\langle\langle\sigma^{2}\rangle_{\sigma|\xi}\xi\rangle_{\xi}
=\langle\langle\sigma\rangle_{\sigma|\xi}\xi^{2}\rangle_{\xi}=0$.
The averages are over the pattern distribution,
Eq.(\ref{1.px}),
and over the conditional distribution, Eq.(\ref{3.ps}):

\begin{equation}
\langle\langle...\rangle_{\sigma|\xi}\rangle_{\xi}=
\sum_{\xi}p(\xi)\sum_{\sigma}p(\sigma|\xi)...
\equiv\langle...\rangle_{\sigma,\xi}.
\label{3.sx}
\end{equation}
Together with the distribution of the patterns,
the conditional probability leads also to the probability

\begin{equation}
p(\sigma)\equiv\sum_{\xi}p(\xi)p(\sigma|\xi)=
q\delta(\sigma^{2}-1)+(1-q)\delta(\sigma).
\label{3.px}
\end{equation}

With the above expressions 
we can calculate the 
$Mutual$ $Information$ $I$\cite{Sh48},\cite{Bl90},
a theoretical information quantity used to measure the average
amount of information that can be received by the user by
observing the symbol (or the signal) at the output of a channel.
We can regard all the dynamical process, 
or rather each time step of it,
as a channel, and write the $I$ as:

\begin{eqnarray}
I[\sigma;\xi]&&=
S[\sigma]-\langle S[\sigma|\xi]\rangle_{\xi};\nonumber\\
S[\sigma]&&
\equiv -\sum_{\sigma}p(\sigma)\ln[p(\sigma)],\nonumber\\
S[\sigma|\xi]&&
\equiv -\sum_{\sigma}p(\sigma|\xi)\ln[p(\sigma|\xi)].
\label{3.Is}
\end{eqnarray}
$S[\sigma]$ and $S[\sigma|\xi]$ are the entropy of the output
and the conditional entropy of the output, respectively.
The quantity $\langle S[\sigma|\xi]\rangle_{\xi}$ is also
called the $equivocation$ $term$ of the $I[\sigma;\xi]$.

For an ANN of homogeneous distributed patterns
the $I$ is not a necessary measure,
because the Hamming distance is enough to quantify the 
quality of the retrieval.
The latter distinguishes well between 
a situation where most of the wrong neurons were turned off 
($d=1$) and another situation where they were flipped ($d=4$).
However, for the low-activity ANN,
$I$ can be a very useful measure,
since the Hamming distance is not so good in distinguishing
the cases where neurons were turned off
from the cases where they were turned on ($d=1$).
This distinction is critical in the sparse coded three-state $NN$,
where $a\ll 1$,
because the inactive states have less information than the
active ones.

For instance, 
let be an ANN with pattern activity $a$
and denote the active (inactive) sites as 
${\cal A}$ (${\cal I}$),
such that $\xi_{\cal A}=\pm 1$ ($\xi_{\cal I}=0$).
Now suppose that all neurons where turned off,
$\sigma_{i}=0$,
then $m=0$, $n=0$ and $q=0$,
so the Hamming distance is $d=a$ and
there is no information transmitted, $I=0$.
If instead of turning off the $A=aN$ active neurons
$\sigma_{\cal A}$,
one had turned on $A$ neurons among the inactive
$\sigma_{\cal I}$,
one get $m=1$, $n=1$ and $q=2a$.
So the Hamming distance is still $d=a$,
but now there is some transmitted information.
It is intuitive that the first kind of errors have erased 
all the meaningful bits,
while the second situation have not affected essentially the code,
and obviously have much less important errors.

The expressions for the entropies defined above are:

\begin{eqnarray}
&&S[\sigma]=
- q\ln{q\over 2} - (1-q)\ln(1-q), \nonumber\\
&&\langle S[\sigma|\xi]\rangle_{\xi}=
a S_{a} + (1-a) S_{1-a},  \nonumber\\
&&S_{a}= 
- {n+m\over 2}\ln{n+m\over 2} 
- {n-m\over 2}\ln{n-m\over 2} \nonumber\\
&&- (1-n)\ln(1-n),  \nonumber\\
&&S_{1-a}=
- s\ln{s\over 2} - (1-s)\ln(1-s).
\label{3.Hs}
\end{eqnarray}

Applying to the second case cited above,
the entropy of the output is 
$S[\sigma]=-2a\ln a-(1-2a)\ln(1-2a)$,
while the equivocation is
$<S[\sigma|\xi]>=
-a[\ln a-\ln 2-\ln(1-a)]-(1-2a)[\ln(1-2a)-\ln(1-a)]$,
so that the mutual information is
$I=-a\ln(2a)-(1-a)\ln(1-a)=S[\xi]-2a\ln(2)$
which is not so smaller than the entropy of the 
original patterns, $S[\xi]$, Eq.(\ref{1.Hm}).
It is easy to understand why we must 
use $I$ for the sparse code case,
instead of the Hamming distance.

\subsection{Derivation of the Hamiltonian}

We search for a Hamiltonian which is symmetric in
any permutations of the patterns $\xi^{\mu}$,
since they are not known during the retrieval process.
This imposes that the retrieval of any pattern $\xi^{\mu}$ 
is week, i.e., $\sigma$ is almost independent of it.
Then obviously the overlap $m^{\mu}\sim 0$.
An expansion of $I$ with $a=1=q$ around $m^{\mu}\sim 0$
yields the Hopfield Hamiltonian.
If afterwards some particular overlap becomes eventually large,
this should be a consequence of the network evolution. 

However, for general $a,q$,
this is not the only quantity which vanishes in this limit.
The variable $\sigma^{2}$ is also almost independent of
$(\xi^{\mu})^{2}$, so that $n^{\mu}\sim q$.
Hence, the parameter 

\begin{equation}
l^{\mu}\equiv\frac{n^{\mu}-q}{1-a}= <\sigma^2\eta^{\mu}>,\,\,
\eta^{\mu}\equiv\frac{{(\xi^{\mu})}^{2}-a}{a(1-a)},
\label{3.lm}
\end{equation}
also vanishes when the states of the neurons and the patterns 
are independent.

We use this fact to look at the information close to the 
non-retrieval regime.
An expansion of the expression for the $I$
around $m^{\mu}=0, l^{\mu}=0$ gives

\begin{equation}
I^{\mu}\approx {1\over 2}{a\over q} {(m^{\mu})}^{2} +
{1\over 2}{a(1-a)\over q(1-q)} {(l^{\mu})}^2.
\label{3.I1}
\end{equation}

Since this expression gives the information for a single
site $i$ of a single pattern $\mu$,
$I(m^{\mu},l^{\mu})\equiv I^{\mu}$,
it should be summed 
$I_{pN}=N\sum_{\mu}I^{\mu}$
to give the total information of the network.
It is natural to associate this quantity with the opposite of the
Hamiltonian, 
because the maximum of the information gives the minimal energy.

We suppose,
as a further simplification of the model,
that the neural activity is of the same order of the pattern
activity, $q \sim a$. 
With this assumption,
$I$ from Eq.(\ref{3.I1}) depends on the same way on
$m^{\mu}$ and $l^{\mu}$. 
Substituting the expressions for these parameters, 
given by the definitions
(\ref{2.mm}),(\ref{2.qN}) and (\ref{2.nN})
($i.e.$, Eqs.(\ref{3.ma}) before the thermodynamic limit),
we obtain the following expression
for the $I$:

\begin{equation}
{\cal H} = -I_{pN} \equiv 
{\cal H}_{1} +  {\cal H}_{2},
\label{3.HI}
\end{equation}
where

\begin{equation}
{\cal H}_{1} = -\frac{1}{2} \sum_{ij} 
J_{ij} \sigma_i \sigma_j
\label{3.H1}
\end{equation}
and

\begin{equation}
{\cal H}_{2} = -\frac{1}{2} \sum_{i,j} 
K_{ij} \sigma_i^2 \sigma_j^2
\label{3.H2}
\end{equation}
are the quadratic and the biquadratic terms,
respectively. 
The above expression for the Hamiltonian, 
obtained from the mutual information close to the 
non-retrieval regime, 
has the same form as of the BEG model\cite{BE71}.
We call our model the BEG Neural Network (BEGNN).

The interactions are randomly distributed,
given by

\begin{equation}
J_{ij}= \frac{1}{a^2 N}
\sum_{\mu=1}^{p}
\xi_{i}^{\mu} \xi_{j}^{\mu},
\label{3.Ji}
\end{equation}
and

\begin{equation}
K_{ij}= \frac{1}{N}
\sum_{\mu=1}^{p}
\eta_{i}^{\mu} \eta_{j}^{\mu} .
\label{3.Ki}
\end{equation}

The first term of the Hamiltonian is the usual Hopfield model
with the Hebbian rule given by Eq.(\ref{3.Ji}).
The second term, 
arising from the term depending on $l^{\mu}$ in Eq.(\ref{3.I1}),
related to the activity-overlap,
is also Hebbian-like, but is associated, 
as will be seen latter, 
with the quadrupolar order of the system.

Note that the Hamiltonian formulation of the problem
is only possible in the case of fully-connected neural network, 
where the interaction matrix is symmetric. 
In the next Section we will present the dynamical formulation 
of the problem, 
which can be applied to the cases 
of asymmetric couplings\cite{DG87}.

As is well known, 
the phase diagram of the usual BEG model is very rich, 
showing different phases, 
depending on the sign and the strength of the biquadratic
coupling constant. 
Without any disorder and for very negative biquadratic
coupling constant, 
a quadrupolar phase, 
related to the quadrupolar moment $<\sigma^2>$ also appear, 
apart of the usual disordered and ferromagnetic phases
\cite{Za95}-\cite{KE99}.
However,
our variables $\xi_i^\mu$ are quenched,
so we have a disordered system.
BEG models with disordered  quadratic coupling
have been recently studied\cite{SN97}-\cite{Br99},
showing  some new phases 
(spin-glass, quadrupolar spin-glass phases, etc),
but, from our knowledge,
no disordered biquadratic BEG model has been
studied up to this date.

\section{Asymptotic Macro-dynamics}

For the  derivation of the asymptotic macro-dynamics 
we will use a naive mean-field (MF) approach using 
the Hamiltonian Eqs.(\ref{3.H1})-(\ref{3.H2}). 
Since the Hamiltonian is quadratic in the overlaps,
we can linearize it, using Gaussian transformation,
to obtain the partition function:

\begin{eqnarray}
&&Z= Tr_{\{\sigma}\}e^{-\beta {\cal H}}= \\
&&\int \prod_{\mu}[
D\Phi(\sqrt{\beta N}m^{\mu}) 
D\Phi(\sqrt{\beta N}l^{\mu})]
\prod_{i}\sum_{\sigma=\pm1,0}e^{\tilde{{\cal H}_i}}
\nonumber ,
\label{4.ZT}
\end{eqnarray}
where $D\Phi(z)\equiv dz e^{-\frac{z^2}{2}}/\sqrt{2\pi}$,
and $\beta=1/T$.
The effective Hamiltonian is

\begin{equation}
\tilde{{\cal H}_i}=
h_i\sigma_i+\theta_i\sigma^2_i,
\label{4.Hi}
\end{equation}
where the local fields are

\begin{eqnarray}
&&h_i=\frac{1}{a}\sum_{\mu}^{p}\xi_i^{\mu}m^{\mu},\,\,
\nonumber\\
&&\theta_i=\sum_{\mu}^{p}\eta_i^{\mu}l^{\mu}.
\label{4.hi}
\end{eqnarray}

After taking the trace over the spin variables,
we apply a saddle point integration 
and use Eq.(\ref{2.FN}) for the thermodynamic limit,
to get the free energy in terms of the parameters $m$, 
$l$ and $q$: 

\begin{equation}
f=-\frac{T}{N}\ln Z= 
\frac{1}{2}(\vec{m}^2+\vec{l}^2)-T<\ln \tilde{Z}>_{\vec{\xi}},
\label{4.fT}
\end{equation}
where the effective partition function is:

\begin{equation}
\tilde{Z}=1+2e^{\beta\theta}\cosh(\beta h).
\label{4.Z1}
\end{equation}
The fields $h,\theta$ are defined in Eq.(\ref{4.hi}),
but the indices $i$ can be dropped out.
The saddle-point equations 
$\partial f/\partial m^{\mu}=0$ and
$\partial f/\partial l^{\mu}=0$.
leads to the following expressions for the 
stationary states:

\begin{eqnarray}
&&m^{\mu}= <\frac{1}{a}\xi^{\mu}\overline{\sigma}>_{\vec{\xi}}, 
\nonumber \\
&&l^{\mu}= <\eta^{\mu}\overline{\sigma^2}>_{\vec{\xi}},
\label{4.ml}
\end{eqnarray}
where the angular brackets mean the average over the patterns, 
and the thermal averages of the states are:

\begin{eqnarray}
&&\overline{\sigma}\equiv F_{\beta}(h,\theta)=
\frac{2e^{\beta\theta}\sinh(\beta h)}{\tilde{Z}},
\nonumber \\
&&\overline{\sigma^2}\equiv G_{\beta}(h,\theta)= 
\frac{2e^{\beta\theta}\cosh(\beta h)}{\tilde{Z}}.
\label{4.sF}
\end{eqnarray}

For zero-temperature, 
the behavior of the averages are:

\begin{eqnarray}
F_{\infty}(h,\theta)=&&
sign(h)\Theta(|h|+\theta),\nonumber\\
G_{\infty}(h,\theta)=&&
\Theta(|h|+\theta),
\label{4.FG}
\end{eqnarray}
where $\Theta(...)$ is the step function.

This result, obtained from the naive MF theory,
can be easily understood if we write the Hamiltonian
in Eqs.(\ref{3.H1},\ref{3.H2}) in the form:

\begin{eqnarray} 
&&{\cal H}=\frac{1}{2}\sum_i\tilde{{\cal H}_i}=
\frac{1}{2}\sum_i(h_i\sigma_i+\theta_i\sigma_i^2);
\nonumber\\
&&h_i\equiv\sum_j J_{ij}\sigma_j, \nonumber\\
&&\theta_i\equiv\sum_{j} K_{ij}\sigma_j^2.
\label{4.Hh}
\end{eqnarray}
So, the deterministic parallel dynamics, 
which leads to the minimization of the Hamiltonian, 
is

\begin{equation} 
\sigma_{i,t+1}=
sign(h_i^t)\Theta(|h_i^t|+\theta_i^t),
\label{4.si}
\end{equation}
where the local fields $h_i^t,\theta_i^t$
(associated to the variables 
$\sigma_{j,t},\sigma_{j,t}^2$ respectively),
are given in the time step $t$.
Such dynamics has the same form as the
zero-temperature function in Eq.(\ref{4.FG}).

Alternatively to the thermodynamic approach, 
in the noise case, 
we can also start from the 
stochastic parallel dynamics\cite{BV94},\cite{BR94}:

\begin{eqnarray} 
p(\sigma_{i,t+1}|\{\sigma_{t}\})=
\exp[\beta\tilde{{\cal H}_i^{t}}]/\tilde{Z},
\label{4.ps}
\end{eqnarray}
where $\tilde{{\cal H}_i^{t}}$ is given by Eq.(\ref{4.Hi})
(in the time step $t$),
and $\tilde{Z}$ by Eq.(\ref{4.Z1}).
Differently from the dynamics for the 
(Q=3)-Ising model\cite{BV94}\cite{BR94},
here the field $\theta=\theta(\{\sigma_j^2\})$ 
in the effective Hamiltonian is a 
function of the states in the previous time steps.
The resulting noise-averaged states coincide with
Eqs.(\ref{4.sF}) in the stationary regime.

Because we are mainly interested on the 
retrieval properties of our network,
we take an initial configuration whose retrieval 
overlaps are only macroscopic of order $O(1)$ 
for a given pattern, 
let say the first one.
We singled out the term $\mu=1$ in the local
fields of Eq.(\ref{4.hi}) in order
to study the retrieval of the first pattern.

Supposing an initial configuration $\{\sigma_{i,t=0}\}$ as 
a collection of IIDRV with zero-mean and variance $q_{t=0}$,
the fields $h_{t=0}$ and $\theta_{t=0}$ in the 
zeroth time step are given by:

\begin{eqnarray}
h_{t=0}=\frac{1}{a}\xi m_{t=0}+\omega_{t=0};\,\, &&
\omega_{t=0}\equiv
\sum_{\nu\geq 2}^{p}\frac{1}{a}\xi^{\nu}m^{\nu}_{t=0}
\nonumber \\
\theta_{t=0}=\eta l_{t=0}+\Omega_{t=0} ;\,\, &&
\Omega_{t=0}\equiv\sum_{\nu\geq 2}^{p}\eta^{\nu}l^{\nu}_{t=0},
\label{4.ht}
\end{eqnarray}
where the indices $\mu=1$ where dropped, 
and the rest of the patterns is regarded as some additive noise.
According to the Central Limit Theorem (CLT),
they are independent Gaussian 
distributed\cite{BV94},\cite{BS94},
with zero mean and variance

\begin{eqnarray}
&&Var[\omega_{t=0}]= 
\frac{1}{a^2}\alpha q_{t=0} \equiv\Delta^2
\nonumber\\
&&Var[\Omega_{t=0}]=
\frac{\Delta^2}{(1-a)^2} .
\label{4.oN}
\end{eqnarray}

Although the dynamics for the parameters $m_t$, $n_t$ and $q_t$
in the first time step is a function of the initial step,
the expression for the noises in the next steps evolves with
time in more complicated way then Eqs.(\ref{4.oN}).
In the extremely diluted synaptic case\cite{DG87}, however,
the first time step describes the dynamics 
for every time step $t$.
From now on we will adopt this limiting case.

Thus, 
in the asymptotic limit $N\to\infty$, 
the expression for the overlap 
$m_{t}=\lim_{N\to\infty}m^{1}_{Nt}$ 
becomes,
after averaging over the pattern $\xi$:

\begin{eqnarray}
\label{4.mt}
&&m_{t+1}= 
<\frac{\xi}{a}\overline{\sigma_t}>_{\vec{\xi}}=\\
&&\int D\Phi(y)\int D\Phi(z)
F_{\beta}(\frac{m_t}{a}+y\Delta_t;
\frac{l_t}{a}+z\frac{\Delta_t}{1-a}),\nonumber
\end{eqnarray}
where the averages over $\omega,\Omega$
on the brackets should be done with the Gaussian distributions, 
Eq(\ref{4.oN}).

The neural activity is the thermodynamic limit of Eq.(\ref{2.qN}),
which reads

\begin{eqnarray}
\label{4.qt}
&&q_{t}=<\overline{\sigma^2_t}>_{\vec{\xi}}= 
a n_{t} + (1-a) s_{t},\\
&&s_{t+1}\equiv \int D\Phi(y)\int D\Phi(z)
G_{\beta}(y\Delta_t;-\frac{l_t}{1-a}+z\frac{\Delta_t}{1-a}).
\nonumber
\end{eqnarray}
Here $s$ is the variable defined in Eq.(\ref{3.ps})  
and the activity-overlap is given by

\begin{eqnarray}
\label{4.nt}
&&n_{t+1}=
<\frac{\xi^2}{a}\overline{\sigma^2_t}>_{\vec{\xi}}=\\
&&\int D\Phi(y)\int D\Phi(z)
G_{\beta}(\frac{m_t}{a}+y\Delta_t;
\frac{l_t}{a}+z\frac{\Delta_t}{1-a}) .\nonumber
\end{eqnarray}

The equation for $l_t$ is obtained using the
definition in Eq.(\ref{3.lm}),
$l_t=(n_t-q_t)/(1-a)$.
It is worth to note that the definitions of the parameters
$m,q,n$ in Eqs.(\ref{3.ma}) are the same as that in
Eqs.(\ref{4.mt}-\ref{4.nt}),
since the average over the conditional probability
$p(\sigma|\xi)$ is equivalent to the average over the noise
due the $p-1$ remaining patterns $\{\vec{\xi}|\xi\}$.
Eqs.(\ref{4.mt}-\ref{4.nt})
describe the macro-dynamics of the diluted BEGNN by 
adapting self-consistently the threshold during the 
time-evolution of the system.
With these equations we can calculate  
the mutual information from Eqs.(\ref{3.Is}-\ref{3.Hs}).

\section{Phase diagram}

In this Section we present some explicit results
for the BEGNN model.
We first calculated the stable fixed-points of the 
Eqs.(\ref{4.mt}-\ref{4.nt})
for the asymptotic $N\to\infty$ network,
and obtained the curves for the order parameters
$m,q,n$ and the {\em information} $i=I\alpha$ 
as a function of the load parameter $\alpha$ for 
two values of the activity $a$ (Fig.1).
For small load ($\alpha<0.2$),
the overlap remains close to $m\sim 1$ and the
neural activity is $q\sim a$. 
When more patterns are stored in the network,
$i$ increases almost linearly,
up to an optimal value, 
$i_{opt}(\alpha_{opt})$,
after which $i$ decreases to zero in $\alpha_{max}$.
The comparison is done with the self-control neural network
($SCNN$) model\cite{DB98},\cite{BD98}. 
It is seen that for small activities ($a=0.3$),
the BEGNN model gives worst results compared with the
SCNN model,
with a smaller value for $i$,
while for $a=0.6$
(close to the uniform distribution of patterns, $a=2/3$),
the BEGNN performs better, 
with an optimal value of the information $i\sim 0.15$,
although it is attained for a smaller value of load,
$\alpha\sim 0.2$.
The reason for this behavior is that the third order parameter
(related to the activity-overlap),
is $l\sim 1$ for the BEGNN (SCNN) 
and $l\ll 1$ for the SCNN (BEGNN)
with $a=0.6$ ($a=0.3$).

The behavior of the order parameters and the $i$ with load,
for the zero-temperature case is presented for three different 
values of the activities (Fig.2).
The initial conditions used where $m_0=10^{-6},l_0=1,q_0=a$,
such that there is almost no initial overlap.
In this case there is always a sharp fall on the 
information for $\alpha$ not so larger than $\alpha_{opt}$. 
We see different behaviors depending on the activities.

The corresponding dynamical phase diagram is drawn in Fig.3. 
Four possible phases are present:
the retrieval R ($m\neq 0,l\neq 0,q\sim a$) 
and M ($m\neq 0,l<0.5m,q\sim a$) phases,
the quadrupolar phase Q ($m=0,l\neq 0,q\sim a$) 
and the zero phase Z ($m=0,l=0,q\sim a$).
The last phase Z,
so called because there is no information transmitted,
is analogue to the Self-sustained (S) activity 
phase of the (Q=3)-Ising ANN\cite{BV94}\cite{BR94},
since the parameter related to the spin-glass order is
$q\neq 0$. 
We have not find any paramagnetic (P) phase,
with all ($m=0,l=0,q=0$) for the BEGNN.
Note that the quadrupolar phase is a quite new phase, 
compared to the other NN models and is a special one for the 
BEGNN-model. 
This phase is also present in the original BEG-model\cite{BE71}, 
as well as in all its generalizations including 
disorder\cite{SN97}-\cite{Br99}.
It is seen in Fig.2, for $a=0.9$, 
where the overlap goes to $m=0$ at $\alpha\sim 0.13$ 
(much before $l$, which goes to zero at $\alpha\sim 0.3$);
this phase corresponds to non-zero information,
although there is no retrieval overlap.
The phase R appears for $a=0.5$, where both $m$ and $l$
are large and so is $i$.
On the other hand,
the phase M is observed for $a=0.1$ where
the parameter $l$ is much smaller than $m$.
The phase transitions from $R$ or $M$ to $Z$ 
are usually sharp.

The behavior of the order parameters and the 
information with the temperature $T$ for fixed 
activity $a=0.5$ is shown on Fig.4. 
We observe an increase of $i$ with the temperature, 
showing an optimal value for $T\sim 0.2$.
Such an improvement of a feeble signal with noise,
similar to the stochastic resonance phenomena,
appears also in other physical systems\cite{GH98}.
A further increase in temperature leads to decreasing 
the information of the model.
We note that this behavior doesn't hold for $a\geq 2/3$,
nevertheless there is still an increase
of the storage capacity $\alpha_{max}$. 
The last result is in agreement with other investigations 
of dynamical activity of real and model neurons, 
where the observed stochastic resonance disappears by 
increasing the amplitude of the external stimulus.

A cut of the phase diagram in the plane 
$T\times\alpha$ for a fixed value of the activity 
$a=0.7$ is shown in Fig.5.  
The dashed line,
which corresponds to the optimal case, 
$i_{opt}(\alpha)$, 
is within either the phase R or Q. 
It is also interesting to observe that there are
two separate Q-phase islands,
for either small temperature $T$ and large load $\alpha$
or large temperature $T$ and small load $\alpha$.
The phase transitions become smoother with the temperature.

Finally on Fig.6 we present the evolution of the information
and of the order parameters with the time $t$,
for a given temperature $T=0.2$ and activity $a=0.7$,
for two values of the load parameter $\alpha$.
As can be seen from this figure, for $\alpha=0.4$, 
which is close to the transition R-M, 
the change to the behavior of the order parameters
needs more time steps than for $\alpha=0.2$.
This is not strange due to the critical slowing down
near the transition.
However, an interesting new fact appears here:
the parameters $l_t$ and $q_t$ have a fast felt down 
to a much smaller value,
after which the network stays a long while with an 
almost zero overlap,
and finally the BEGNN is able to retrieve quite well the pattern.
For instance, for $\alpha=0.2$,
$l$ falls to $l\sim 0.6$ and $m$ stays near $m\sim 0$
during the first $t\sim 20$ time steps,
then they jump up to $l\sim 0.8$, $m\sim 0.9$,
which means the memory pattern was (partially) attained.
This result,
caused by the instability of the Z-phase in this region,
makes the BEGNN capacity much larger than that of the usual
Hopfield model,
in all its versions so far as we know.

The behavior of the continuous phase transitions can be 
analytically studied within the mean-field approximation 
by expanding Eqs.(\ref{4.mt}-\ref{4.nt}) for small
values of the order parameters.
A standard calculation, for example, 
for the transition line  QZ ($m=0, l<<1$) leads to the 
following expression:

\begin{eqnarray}
\label{trQSl}
&&l^{QZ} = 
\beta T_{c}^{QZ}l + \nonumber\\
&& \left<e^{\beta \Omega}\cosh{\beta\omega}
\frac{1-2 e^{\beta \Omega} \cosh{\beta\omega}}
{(1+2 e^{\beta \Omega} \cosh{\beta \omega})^3}
\right>_{\Omega,\omega}
\frac{(1-2a)\beta^2}{a^2 \hat{a}^2}l^2 
,
\end{eqnarray}
where the transition temperature between the phases Q and Z is:

\begin{equation}
\label{trQSt}
T_{c}^{QZ} = \frac{1}{a \hat{a}} 
\left<\frac{2 e^{\beta\Omega} \cosh{\beta\omega}}
{(1+2 e^{\beta\Omega} 
\cosh{\beta\omega})^2}\right>_{\Omega,\omega}
\end{equation}
with
\begin{equation}
\label{trQSq}
q^{QZ} = \left<\frac{2 e^{\beta\Omega} \cosh{\beta\omega}}
{1+2 e^{\beta\Omega} \cosh{\beta\omega}}\right>_{\Omega,\omega} 
+ O(l^2).
\end{equation}   

Expanding the above expressions for small value of the load rate
and large temperatures, $\beta\sqrt{\alpha}\ll 1$,
and calculating the averages over the noise up to the leading terms, 
one obtains the following equation for the transition line:

\begin{equation}
\label{T_c}
T_c= \frac{2}{9 a \hat{a}} - 
\frac{1}{2}\frac{(1+\hat{a}^2)}{a \hat{a}}\alpha .
\end{equation}
The last expression for $T_c$ is in qualitative agreement with 
the previous results shown on Fig.5.

Regarding the equation for the order parameter $l$, 
one can verify that in
leading order:

\begin{equation}
l^{QZ}=\beta T_{c}^{QZ}l - 
\frac{1}{27}\frac{(1-2a)}{a^2\hat{a}^2}\beta^2 l^2 +
O(l^3, \alpha l^2).
\end{equation}
By use of Eq.(\ref{T_c}), 
it is seen that the quadratic term of the above expansion 
changes sign when the activity $a=0.5$, 
thus defining a tricritical line between the transition
of second order ($a>0.5$) and of first order ($a<0.5$). 
Note that similar tricritical behavior has been described 
also in the other versions
of the BEG model \cite{Za95}-\cite{Br99}.
Similar analysis can be also performed for the other 
continuous transition between the different phases.

\section{Conclusions}

In this paper we proposed a BEG-like Hamiltonian
for a ternary neural network, 
which couplings arise from an expansion of its 
mean-field mutual information, 
$I$\cite{BD98},
resulting in a system evolving with a 
self-consistently adapting threshold.
The stationary and dynamical equations for this model
were obtained 
as functions of three order parameters,
the overlap $m$, the neural activity $q$,
and the activity-overlap $n$.
Their solutions were explicitly calculated as
functions of the variables: 
the pattern activity $a$, 
the load $\alpha$ and the temperature $T$.

When the activity is near $a=2/3$,
corresponding to the uniform ternary patterns,
the BEGNN improves the information,
compared with a previous SCNN model\cite{DB98},\cite{BD98}.
Improvement of the information content by increasing the noise,
effect similar to the stochastic resonance, 
is also observed for activities $a<2/3$.

There are four possible phases for the BEGNN,
which were displayed in phase diagrams $a\times\alpha$
and $T\times\alpha$.
In particular, 
a quadrupolar phase, Q,
with $m=0,l\sim 1$,
holds whenever the activity is large enough.
This phase,
known in the BEG literature,
but new in an ANN context,
carries out some nonzero information about the patterns
even without any overlap $m$.

As the main result we obtained that,
while the phase Z is not stable in a large range
of the variables,
the basin of attraction of the retrieval phase
is increased with respect to the usual ternary
neural network models.
States with initial conditions having very small overlap
flow to final states with large overlap.

We believe that the BEGNN has a quite large
range of applications for real systems.
We also think that this way to obtain an Hamiltonian starting
from a mean-field calculation of $I$,
which yields an almost optimal retrieval dynamics, 
can be generalized to other spin systems,
as the $Q$-Ising with $Q>3$ or the Potts models, for instances.
Such method, based on the maximization of the entropy,
can be an universal approach to information systems.

Then, 
we expect that the same improvement should happen for 
analogue neurons and for networks of binary {\it synapses}.
It would be also interesting to investigate the case
of local field for multi-neuron synapses,
which comes up from higher order terms in the expansion 
of the mutual information,
such that a better use of a network with fixed size is expected. 
 
\vskip0.5cm
{\bf Acknowledgments}

We thank the Workshop on 
"Statistical Mechanics of Neural Networks", 
Max-Planck Institute, Dresden'99,  for useful discussions.
E.K. is financially supported by Spanish DGES grant
PB97-0076 and partly by contract F-608 with Bulgarian
Scientific Foundation.



\vspace{-1cm}
\begin{figure}[t]
\begin{center}
\epsfysize=8cm
\leavevmode
\epsfbox[1 1 700 700]{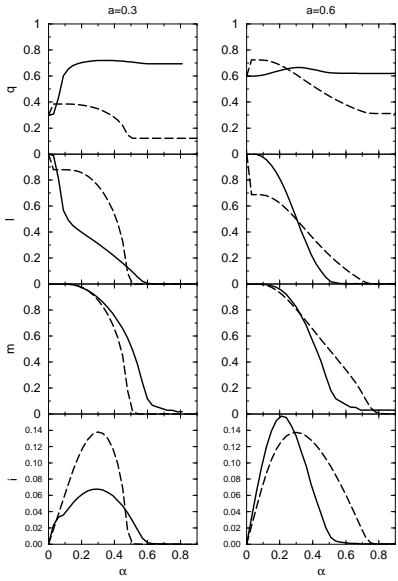}
\end{center}
\caption{ {\em 
The information $i=I\alpha$ and the order parameters
$m,l,q$ against $\alpha$ with activities
$a=0.3$ (left) and $a=0.6$ (right).
The temperature $T=0$ and the initial conditions are
$m_0=1$, $l_0=1$, $q_0=a$.
The continuous line is for the BEGNN while the
dashed line is for the SCNN. } }
\label{I,tA}
\end{figure}

\vspace{-1cm}
\begin{figure}[b]
\begin{center}
\epsfysize=8cm
\leavevmode
\epsfbox[1 1 700 700]{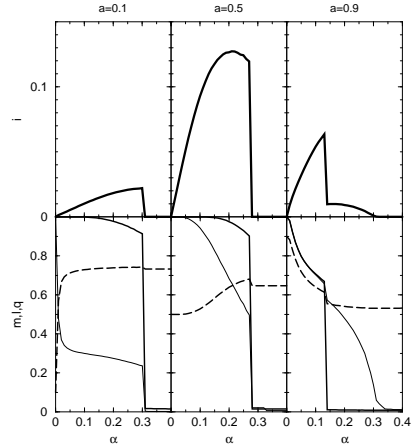}
\end{center}
\caption{ {\em 
The information $i$ and the order parameters
$m$ (solid line),
$l$ (thin line) and $q$ (dashed line)
against $\alpha$ with activities 
$a=0.1$ (left), $a=0.5$ (center) and $a=0.9$ (right);
The temperature $T=0$ and the initial conditions are
$m_0=10^{-6}$, $l_0=1$, $q_0=a$. } }
\label{m,tA}
\end{figure}

\vspace{-1cm}
\begin{figure}[b]
\begin{center}
\epsfysize=8cm
\leavevmode
\epsfbox[1 1 700 700]{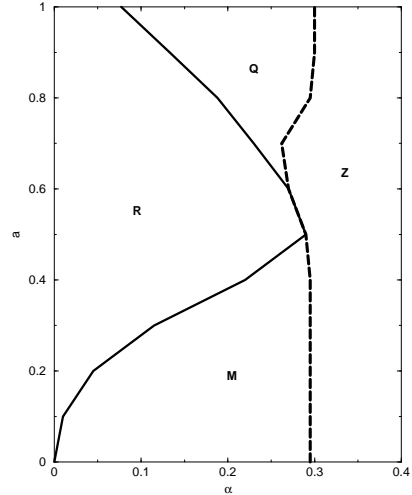}
\end{center}
\caption{ {\em 
The dynamical phase diagram $\alpha\times a$,
for $T=0$ with initial conditions 
$m_0=10^{-6}$, $l_0=1$, $q_0=a$.
The different phases are explained in the text.} }
\label{m0,a}
\end{figure}

\vspace{-1cm}
\begin{figure}[h]
\begin{center}
\epsfysize=9cm
\leavevmode
\epsfbox[1 1 700 700]{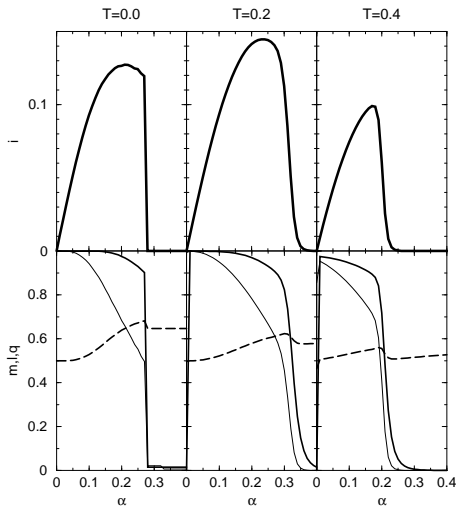}
\end{center}
\caption{ {\em 
The information $i$ and the order parameters
$m$ (solid line),
$l$ (thin line) and $q$ (dashed line)
against $\alpha$ with temperatures
$T=0.0$ (left), $T=0.2$ (center) and $T=0.4$ (right);
The activity $a=0.5$ and the initial conditions are
$m_0=10^{-6}$, $l_0=1$, $q_0=a$. } }
\label{Ii,a}
\end{figure}

\vspace{-4cm}
\begin{figure}[h]
\begin{center}
\epsfysize=9cm
\leavevmode
\epsfbox[1 1 700 700]{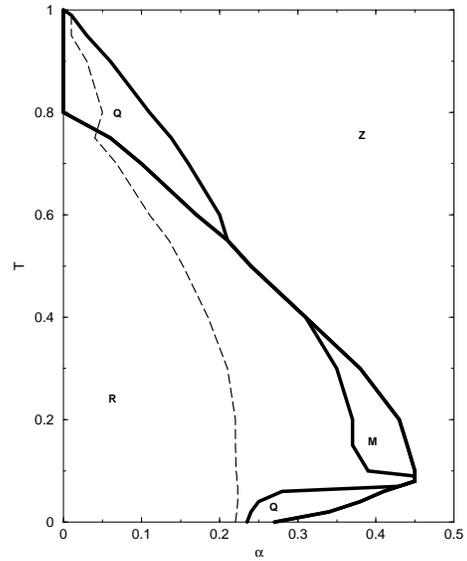}
\end{center}
\caption{ {\em 
The dynamical phase diagram $\alpha\times T$,
for $a=0.7$ with initial conditions 
$m_0=10^{-6}$, $l_0=1$, $q_0=a$. 
The dashed line corresponds to the optimal information. } }
\label{ia,A}
\end{figure}

\vspace{-1cm}
\begin{figure}[h]
\begin{center}
\epsfysize=9cm
\leavevmode
\epsfbox[1 1 700 700]{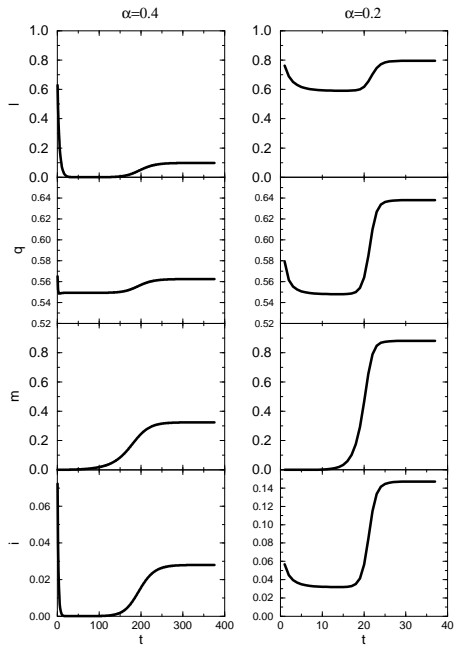}
\end{center}
\caption{ {\em 
The information $i$ 
and the order parameters $m,l,q$,
against the time $t$ for temperature $T=0.2$
and activity $a=0.7$,
with $\alpha=0.4$ (left) and $\alpha=0.2$ (right).
The initial conditions are $m_0=10^{-6}$, $l_0=1$, $q_0=a$.
} }
\label{im,M}
\end{figure}



\end{document}